\documentclass[prl,superscriptaddress,twocolumn,showpacs]{revtex4}
\usepackage{amsmath,graphicx}

\begin{document}
\title{AC Josephson Effect and Resonant Cooper Pair Tunneling Emission of a Single Cooper Pair Transistor}
\author{P.-M. Billangeon}
\affiliation{Laboratoire de Physique des Solides, University Paris-Sud, CNRS, UMR 8502, F-91405 Orsay Cedex, France.}
\author{F. Pierre}
\affiliation{Laboratoire de Photonique et Nanostructures, CNRS, UPR 20, F-91460 Marcoussis, France.}
\author{H. Bouchiat}
\affiliation{Laboratoire de Physique des Solides, University Paris-Sud, CNRS, UMR 8502, F-91405 Orsay Cedex, France.}
\author{R. Deblock}
\affiliation{Laboratoire de Physique des Solides, University Paris-Sud, CNRS, UMR 8502, F-91405 Orsay Cedex, France.}

\begin{abstract}
We measure the high-frequency emission of a single Cooper pair transistor (SCPT) in the regime where transport is only due to tunneling of Cooper pairs. This is achieved by coupling on-chip the SCPT to a superconductor-insulator-superconductor junction and by measuring the photon assisted tunneling current of quasiparticles across the junction. This technique allows a direct detection of the AC Josephson effect of the SCPT and provides evidence of Landau-Zener transitions for proper gate voltage. The emission in the regime of resonant Cooper pair tunneling is also investigated. It is interpreted in terms of transitions between charge states coupled by the Josephson effect. 
\end{abstract}

\maketitle

One of the most striking consequence of the macroscopic quantum coherence of the superconducting state is the Josephson effect which takes place when two superconductors are connected via a non-superconducting material (insulator or metal).  It leads to a supercurrent at zero bias, and an AC current when the junction is voltage biased \cite{tinkham96,barone82}. In this work we investigate experimentally how the AC Josephson emission is modified when the Josephson effect competes with charging effects. To do so, we consider a single Cooper pair transistor (SCPT) constituted by two small junctions separated by a superconducting island, which energy can be tuned by a nearby electrostatic gate \cite{Averin91}. Because of the interplay between Josephson coupling and charging effects the SCPT exhibits peculiar electronic transport properties extensively studied over the past 15 years \cite{Averin91,Fulton89,Averin89,Geerligs90,Joyez94,JoyezPhD}. However AC Josephson effect was less investigated. To characterize it one can irradiate the SCPT with high frequencies (HFs) and observe Shapiro steps, which result from the locking of the superconducting phase dynamics on the frequency of the irradiating signal \cite{barone82,JoyezPhD}. However these steps may be hard to distinguish from other features of the SCPT. We use another technique to measure directly the AC Josephson effect~: we couple the SCPT on-chip to a HF detector, a superconductor-insulator-superconductor (SIS) junction and measure the photo-assisted tunneling (PAT) quasiparticle current due to the emission of the SCPT \cite{deblock03,billangeon06}. 

\begin{figure}
	\begin{center}
		\includegraphics[width=7.5cm]{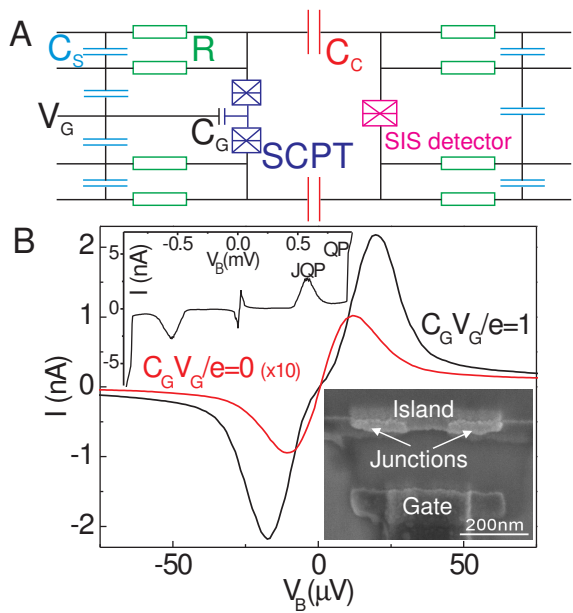}
	\end{center}
	\caption{A : Schematic picture of the SCPT coupled to a Josephson junction and the on-chip circuit ($R=750$ $\Omega$, $C_C \approx 750$ fF). B : Josephson branch of the SCPT at $C_G V_G/e=1$ and $0$ (multiplied by 10). Upper inset : I(V) characteristic of the SCPT at high bias showing the Josephson-Quasiparticle (JQP) peak and the quasiparticle (QP) tunneling. Lower inset : SEM picture of the SCPT.}
	\label{fig:figure1}
\end{figure}

The device probed in this experiment at 90 mK is a SCPT (normal state resistance 48.5 k$\Omega$) coupled capacitively to a small SIS junction (estimated capacitance 1 fF, normal state resistance $R_T=25$ k$\Omega$). Both structures are made in aluminum (superconducting gap $\Delta$ = 210 $\mu$eV) and embedded in an on-chip environment constituted by resistances (8 Pt wires, $R$ = 750 $\Omega$, length = 40 $\mu$m, width = 750 nm, thickness = 15 nm) and capacitances (estimated value $C_C \approx 750$ fF, size~: 23$\times$25 $\mu$m$^2$, insulator~: 65 nm of Al$_2$O$_3$) designed to provide a good HF coupling between the two devices (Fig. \ref{fig:figure1}A). The SIS junction has a SQUID geometry in order to minimize its critical current with a magnetic flux.

We first present transport measurements performed on the SCPT. On Fig. \ref{fig:figure1}B the $I(V_B)$ characteristic for low bias voltage of the SCPT at two gate voltages is shown. At both values, the SCPT shows a Josephson branch which extends to finite voltages as commonly seen for SCPT \cite{Kycia01,Lu02,Lotkhov03} (and Josephson junction \cite{Averin01,Kuzmin91,Ingold92,Ingold99}) embedded in a dissipative environment. The gate dependence of the Josephson branch is 2e periodic, as expected from the hamiltonian of the SCPT :
\begin{eqnarray}
	H &=& \sum_n \left[ E_C (n-C_G V_G/e)^2 |n\rangle \langle n| \right. \nonumber \\ 
	&-& \left. E_J \cos(\delta/2) \left(|n\rangle \langle n+2| + |n+2\rangle \langle n|\right) \right]+ H_S \nonumber
	\label{H}
\end{eqnarray}
with $E_C=e^2/2 C_\Sigma$ the charging energy ($C_\Sigma$ is the total capacitance of the island), $E_J$ the Josephson energy of each junction, $\delta$ the superconducting phase difference between the reservoirs and $|n\rangle$ the state with $n$ electrons on the island. $H_S$ describes a superconducting metal by the BCS theory and favors paired electrons on the island \cite{Tuominen92}, leading to the 2e periodicity, if the superconducting gap is bigger than $E_C$ \cite{JoyezPhD}. The transport measurement of the SCPT allows to determine its charging energy ($E_C=65 \mu$eV) and Josephson energy ($E_J=28 \mu$eV).

\begin{figure}
	\begin{center}
		\includegraphics[width=7.5cm]{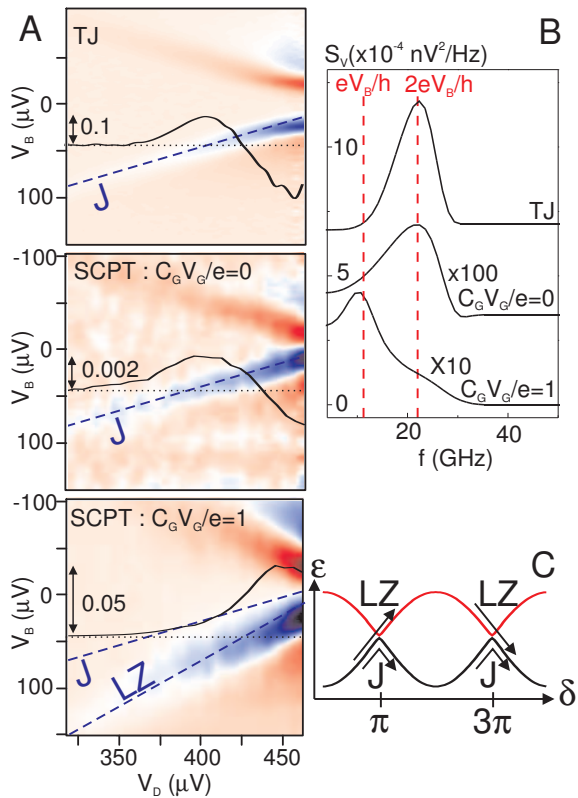}
	\end{center}
	\caption{A) Derivative of the PAT current of the detector versus $V_B$ of the SCPT ($dI_{PAT}/dV_B$) (or a Josephson junction for the upper panel, noted TJ) for $C_G V_G/e=0$ and $1$ (lower panel). The dashed line J (LZ) indicates the slope of the Josephson (Landau-Zener) effect. The solid curves are $dI_{PAT}/dV_B$ at $V_B=45 \mu$V, with the scale in pA/$\mu$V indicated by the arrow on the curve and the zero line shown by the dotted line. B) Voltage power spectrum of the AC Josephson effect extracted from the curves shown on A for a Josephson junction (TJ), and for the SCPT at $C_G V_G/e=0$ and $1$. The curves are shifted for clarity. C) Sketch of the phase dependence of the two lowest energy levels of the SCPT at $C_G V_G/e=1$ and illustration of the Landau-Zener (LZ) and Josephson (J) effect.}
	\label{fig:figure2}
\end{figure}

We now turn to the detection of the AC Josephson effect. For a DC voltage biased junction it results from the evolution of the superconducting phase difference $\delta$ across the junction according to the Josephson relation $d \delta/dt = 2e V_B/\hbar$, with $V_B$ the DC voltage bias. Due to the periodic current-phase relation of the junction, this leads to an oscillating current. Since the SCPT can be considered at low voltage as a Josephson junction with a gate-dependent critical current, it should exhibit a gate dependent AC Josephson effect. To detect it the PAT current through the SIS junction is measured at $C_G V_G/e=0$ and $1$, versus bias voltage $V_B$ and detector voltage $V_D$ (Fig. \ref{fig:figure2}A, two lower panels). To improve sensitivity we modulate $V_B$ and monitor the modulated part of the PAT current $\partial I_{PAT}(V_D)/\partial V_{B}$ with a lock-in technique. The same type of measurement is shown for a small Josephson junction (Fig. \ref{fig:figure2}A, upper panel) \cite{billangeon06}. When biased at $V_D$ the detector is sensitive to photons of energy higher than $eV_D-2\Delta$. This relation between $V_D$ and the energy of detected photons allows a frequency resolved detection. The signature of the AC Josephson effect is then a peak, followed by a dip, on $\partial I_{PAT}(V_D)/\partial V_{B}$ at a detector voltage $V_D$ corresponding to the Josephson frequency $\nu_J$ : $eV_D-2\Delta=h \nu_J$. Since $\nu_J$ depends linearly on the source voltage ($h \nu_J = 2eV_B$ for a Josephson junction), the position of the peak in $\partial I_{PAT}(V_D)/\partial V_{B}$ versus $V_D$ varies linearly with $V_B$, with a slope dependent on the relation between the source voltage and the Josephson frequency. We find that this slope is the same for a Josephson junction and for the SCPT at $C_G V_G/e=0$, and corresponds to the relation expected from the Josephson relation (dashed line of Fig. \ref{fig:figure2}). However for $C_G V_G/e=1$ this slope is divided by a factor 2, as if $h \nu_J = e V_B$. We attribute this factor 2 to the proximity of ground and first excited states which favors Landau-Zener (LZ) transitions. Indeed for $C_G V_G/e=1$  the ground state and the first excited state are nearly degenerate at $\delta=\pi$ modulo $2\pi$ (Fig. \ref{fig:figure2}C). During the phase evolution of the SCPT, the system can there either stay in the ground state (J arrow of Fig. \ref{fig:figure2}C) or go into the excited state (LZ arrow) \cite{junctions}. If this latter transition always happens the effective periodicity of the energy-phase relation, and thus current-phase relation, is doubled \cite{JoyezPhD}, leading to the observed doubling of the Josephson period. This requires three conditions~: first a high probability for LZ transition \cite{junctions}, second a relaxation time longer than the period of the Josephson effect, and third a relaxation time shorter than the time needed to equalize the population of the ground and first excited state during the phase evolution. This implies relaxation time of the order of few nanoseconds, in agreement with relaxation induced by the electromagnetic environment \cite{Ithier05}. Note that poisoning effect cannot lead to the doubling of the Josephson period close to $C_G V_G/e=1$. Without LZ effect, close to $C_G V_G/e=1$, one should expect a strongly non-harmonic current-phase relation. In our case this is hidden by the LZ effect, which restores a sinusoidal current-phase relation.
 
Using a numerical deconvolution method \cite{deblock03}, we extract from the data taken at $V_B=45 \mu$V the voltage power spectrum $S_V(\omega,V_B)$ (Fig. \ref{fig:figure2}B) across the detector leading to the measured PAT current (Fig. \ref{fig:figure2}A). $S_V(\omega,V_B)$ exhibits peaks at the expected frequency. For $C_G V_G/e=1$ we find a peak at $e V_B/h$, attributed to the LZ transitions, but also an extension of the spectrum around $2 e V_B/h$. This might be due to incomplete LZ effect. Due to the rather broad $IV$ characteristic of the detector, those spectra have to be taken cautiously at low frequencies and for a precise determination of the width of the Josephson emission.  We find a rather large emission width around 8 GHz (corresponding to 30 $\mu$V), which is consistent with the width of the Josephson branch measured in DC (Fig. \ref{fig:figure1}B). This seems to indicate that the electromagnetic environment acts similarly on the DC and AC Josephson effect. To analyse the data we assume that the area of the Josephson peak is given by the voltage fluctuation induced by the source and thus proportional to $Z^2 I_C^2$, with $Z$ the transimpedance of the circuit (ratio of the AC voltage at the detector divided by the AC current of the source) and $I_C$ the critical current of the source. $Z$ was measured for the same environment \cite{billangeon06}. We find that the critical current at $C_G V_G/e=0$ is 1.9 nA, compared with the theoretical value 1.45 nA \cite{CalculIc} and 6.5 nA at $C_G V_G/e=1$, compared with an expected value of 6.8 nA. To conclude this part, our direct detection of the emission of the SCPT demonstrates a gate-dependent AC Josephson effect, not only in amplitude but also in frequency due to Landau-Zener transitions.

\begin{figure}
	\begin{center}
		\includegraphics[width=7.5cm]{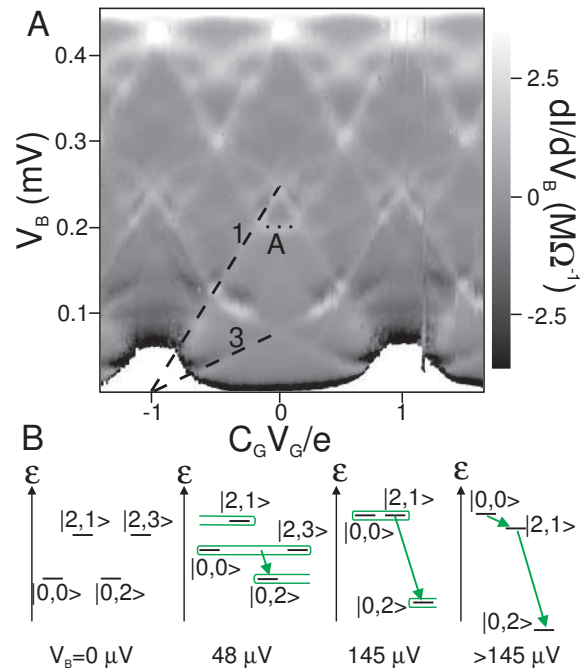}
	\end{center}
	\caption{A) Differential conductance of the SCPT. The resonances "1" and "3" (see text) are indicated on the plot. B) Energy levels of the SCPT at different bias voltage. The arrows indicate transitions between states resulting from the coupling by the Josephson effect of charge states.The green rectangles represent degenerate states coupled by the Josephson term.}
	\label{fig:figure3}
\end{figure}

Besides the DC Josephson peak, the differential conductance of a SCPT exhibits peaks at finite bias \cite{Maasen91,Haviland94}. Hereafter we focus on the region where the source voltage $V_B$ of the SCPT is smaller than $2 \Delta +E_C$, with only tunneling of Cooper pair (CP). For appropriate values of $V_B$ and $V_G$, one observes a sharp increase of the DC current associated to transitions between the CP states of the system leading to peaks of differential conductance $\partial I/ \partial V_B$ (Fig. \ref{fig:figure3}A). The bias voltage has two effects on the SCPT emission : it induces an evolution of the superconducting phase and modifies the energy of the charge states. To describe this last effect it is convenient to consider states with $n$ electrons on the island and $k$ electrons having passed through the SCPT, noted $|n,k\rangle$ \cite{JoyezPhD}. The energy of the state $|n,k\rangle$ in the presence of an applied voltage $V_B$ is changed by $-k e V_B$. It is then possible to draw the diagram of energy levels at different bias voltage for a given gate voltage (Fig. \ref{fig:figure3}B for $C_G V_G/e=0.45$) and deduce the expected transitions and resonance. In this formalism, the AC Josephson effect is a transition induced by the Josephson coupling between levels like $|0,0\rangle$ and $|0,2\rangle$. When $V_B$ is such that the energy of state $|0,0\rangle$ is resonant with another state (e.g., in Fig. \ref{fig:figure3}B, $|2,3\rangle$ at $V_B=48 \mu$V (resonance ``3'') or $|2,1\rangle$ at $145 \mu$V (resonance ``1'')) high order Josephson terms lead to an increase of differential conductance $\partial I/ \partial V_B$ (line ``1'' and ``3'' on Fig. \ref{fig:figure3}A). This resonant Cooper pair tunneling (RCPT) is 2e-periodic, as expected from the hamiltonian of the system, except at high voltage ($V_B > 150 \mu$V) where poisoning (\textit{i.e.} presence of non-paired electrons on the SCPT) restores an e-periodicity.

\begin{figure}
	\begin{center}
		\includegraphics[width=7.5cm]{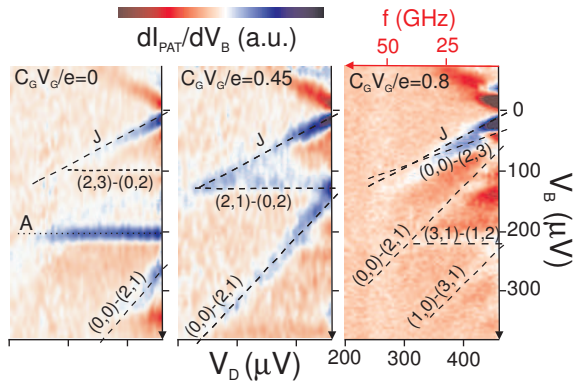}
	\end{center}
	\caption{Derivative of the PAT current of the detector versus bias voltage $V_B$ of the SCPT for different gate voltage, indicated on each plot. The dashed lines indicate the transitions predicted by the theory of RCPT (see text). The dashed line noted J correspond to AC Josephson effect. The color scale is not the same for all plots.}
	\label{fig:figure4}
\end{figure}

Since RCPT implies transitions between charge states induced by the Josephson coupling, it may lead to HF emission \cite{Naaman06}. To characterize this emission, we perform the same type of measurement as for the AC Josephson effect on a wider range of bias and gate voltage (Fig. \ref{fig:figure4}). Beside the already mentioned AC Josephson effect (``J'' dashed line) the SCPT emission at a given source voltage presents peaks at certain frequencies leading to features on the PAT current for particular values of detector bias. We relate them to transitions between quantum states of the system. The energies of these transitions are calculated using the formalism presented before. RCPT happens when $V_B$ is high enough to allow the tunneling of CP, as illustrated on Fig. \ref{fig:figure3}B. When such a tunneling between states $|n,k\rangle$ and $|n,k+2q\rangle$, involving an intermediate state $|n',k'\rangle$, is permitted two emission processes happen sequentially. First the emission of a photon of energy $(k'-k) e (V_B-V_0)$ corresponding to the difference in energy between $|n,k\rangle$ and $|n',k'\rangle$, with $V_0$ the onset voltage where the transition $|n,k\rangle \rightarrow |n',k'\rangle$ happens (\textit{e.g.} transition $|0,0\rangle-|2,1\rangle$ indicated by the dashed line $(0,0)-(2,1)$ on Fig. \ref{fig:figure4}). Second, when $V_B>V_0$, there is the emission of a photon corresponding to the transition from $|n',k'\rangle$ to $|n,k+2q\rangle$, which energy is $2qeV_B-(k'-k)e(V_B-V_0)$. Since we measure the signal derivative of the PAT current versus $V_B$, this transition appears essentially as a step versus $V_B$ at $V_B=V_0$, with an extension in frequency up to $2qeV_B/h$ (\textit{e.g.} the horizontal line $(2,1)-(0,2)$ on Fig. \ref{fig:figure4}). Close to $C_G V_G/e=1$ all the RCPT resonances collapse into the Josephson peak (Fig. \ref{fig:figure3}A). It is then more complicated to separate the different phenomena at small voltage. Thus for $C_G V_G/e=0.8$ the AC Josephson effect (line J) is mixed with a RCPT resonance (line (0,0)-(2,3)). This prevents us to measure accurately the transition to Landau-Zener effect versus gate voltage.

Measurements of high-frequency emission can also provide information on processes not expected from the RCPT theory. As measured on the differential conductance of the SCPT, poisoning leads to an e-periodicity at sufficiently high voltage. We then have to consider the charge states with an odd number of electrons on the island. This is why for $C_G V_G/e = 0.8$, we have signature of transitions between states $|1,0\rangle$ and $|3,1\rangle$ (line (1,0)-(3,1)) and between $|3,1\rangle$ and $|1,2\rangle$ ((line (3,1)-(1,2)). Another process is the feature A (Fig. \ref{fig:figure3}A) at $V_B =205 \mu$V. It exhibits a HF emission at 50 GHz (Fig. 4, $C_G V_G/e=0$). An explanation for this can be that, due to poisoning at $V_B=205 \mu$V, the energy levels with one quasiparticle can be populated. For $C_G V_G/e=0$ the charge state $|1\rangle$ and $|-1\rangle$ are degenerate and coupled by the Josephson effect. This leads to LZ transitions and HF emission at $e V_B/h=50$ GHz, which is indeed the typical frequency of the feature A. This LZ effect for the poisoned SCPT takes place on a region of width 0.3 around $C_G V_G/e=0$. 

In conclusion we have performed a direct detection of the HF emission of a SCPT by coupling it on-chip to a SIS detector. This demonstrates an AC Josephson effect which is gate dependent in amplitude but also in frequency, due to Landau-Zener effect. We have also detected the emission in the RCPT regime, and interpreted the observed peaks as signatures of transitions between charge states coupled by the Josephson effect.


\begin{thebibliography}{100}

\bibitem{tinkham96}
M. Tinkham, \textit{Introduction to Superconductivity} (McGraw Hill, second edition, 1996).

\bibitem{barone82}
A. Barone and G. Paterno, \textit{Physics and Applications of the Josephson effect} (Wiley-Interscience, New-Yok, 1982).

\bibitem{Averin91} D.V. Averin and K.K. Likharev, in Mesoscopic Phenomena in
Solids, edited by B.L. Altshuler \textit{et al.} (Elsevier, Amsterdam,
1991).

\bibitem{Fulton89}
T.A. Fulton \textit{et al.}, Phys. Rev. Lett. \textbf{63}, 1307 (1989). 

\bibitem{Averin89}
D.V. Averin and V.Ya. Aleshkin, JETP Lett. \textbf{50}, 367 (1989).

\bibitem{Geerligs90}
L.J. Geerligs, V.F. Anderegg, J. Romijn and J.E. Mooij, Phys. Rev. Lett. \textbf{65}, 377 (1990).

\bibitem{Joyez94}
P. Joyez \textit{et al.}, Phys. Rev. Lett. \textbf{72}, 2458 (1994). 

\bibitem{JoyezPhD}
P. Joyez, PhD thesis, Paris 6 University,1995. 

\bibitem{deblock03}
R. Deblock, E. Onac, L. Gurevich and L.P. Kouwenhoven, Science \textbf{301}, pp 203-206 (2003).

\bibitem{billangeon06}
P.-M. Billangeon, F. Pierre, H. Bouchiat and R. Deblock, Phys. Rev. Lett. \textbf{96}, 136804 (2006).

\bibitem{Kycia01}
J.B. Kycia \textit{et al.}, Phys. Rev. Lett. \textbf{87}, 017002 (2001).

\bibitem{Lu02}
W. Lu, A.J. Rimberg and K.D. Maranowski, Appl. Phys. Lett. \textbf{81}, 4976 (2002).

\bibitem{Lotkhov03}
S.V. Lotkhov, S.A. Bogoslovsky, A.B. Zorin and J. Niemeyer, Phys. Rev. Lett. \textbf{91}, 197002 (2003).

\bibitem{Averin01}
D.V. Averin, Yu.V. Nazarov and A.A. Odintsov, Physica \textbf{165B \& 166B}, 945 (1990).

\bibitem{Kuzmin91}
L.S. Kuzmin \textit{et al.}, Phys. Rev. Lett. \textbf{67}, 1161 (1991).

\bibitem{Ingold92}
G.-L. Ingold and Yu.V. Nazarov, in \textit{Single-Charge Tunneling}, edited by H. Grabert and M.H. Devoret (Plenum, New-York, 1992).

\bibitem{Ingold99}
G.-L. Ingold and H. Grabert, Phys. Rev. Lett. \textbf{83}, 3721 (1999).

\bibitem{junctions}
From measurements on different junctions, we estimate the asymmetry of tunnel junctions of the SCPT to be less than 10 \%. This leads to a calculated gap in energy at $C_G V_G/e=1$ and $\delta=\pi$ of less than $\epsilon=$2.8 $\mu$eV. The probability of Zener transition $P_Z=\exp(-\pi \epsilon^2/2 E_J eV_B)$ \cite{Mullen88} is then higher than 0.95 at $V_B=10 \mu$V and 0.99 at 45 $\mu$V. 

\bibitem{Mullen88}
K. Mullen, Y. Gefen and E. Ben Jacob, Physica B \textbf{152}, 172 (1988).

\bibitem{Ithier05}
G. Ithier \textit{et al.}, Phys. Rev. B. \textbf{72}, 134519 (2005).

\bibitem{Tuominen92}
M.T. Tuominen, J.M. Hergenrother, T.S. Tighe and M. Tinkham, Phys. Rev. Lett. \textbf{69}, 1997 (1992).

\bibitem{CalculIc}
This is deduced from the calculated energy levels and the current-energy relation$I(\delta)= 2e/\hbar \, \partial E(\delta)/\partial \delta$.

\bibitem{Maasen91}
A. Maassen van den Brink, G. Sch\"on, and L.J. Geerligs,  Phys. Rev. Lett. \textbf{67}, 3030 (1991).

\bibitem{Haviland94}
D.B. Haviland \textit{et al.},  Phys. Rev. Lett. \textbf{73}, 1541 (1994).

\bibitem{Naaman06}
O. Naaman and J. Aumentado, cond-mat/0609491.

\end{thebibliography}
\end{document}